\documentclass[12pt]{article}

\usepackage{amsmath,amsfonts,amssymb}
\usepackage{graphicx}
\usepackage{psfrag}
\usepackage{enumerate}
\usepackage{mathrsfs}
\usepackage{cancel}
\usepackage{multirow}
\usepackage{url}

\makeatletter\renewcommand\section{\@startsection {section}{1}{\z@}%
                                   {-3.5ex \@plus -1ex \@minus -.2ex}%nn
                                   {2.3ex \@plus.2ex}%
                                   {\normalfont\large\bfseries}}
\renewcommand\subsection{\@startsection{subsection}{2}{\z@}%
                                     {-3.25ex\@plus -1ex \@minus -.2ex}%
                                     {1.5ex \@plus .2ex}%
                                     {\normalfont\bfseries}}

\newcommand{\be}{\begin{equation}}
\newcommand{\ee}{\end{equation}}
\newcommand{\beq}{\begin{eqnarray}}
\newcommand{\eeq}{\end{eqnarray}}

\def\[{\left [}
\def\]{\right ]}
\def\({\left (}
\def\){\right )}

\def\r2{\sqrt{2}}

\newcommand{\bbibitem}[1]{\bibitem{#1}\marginpar{#1}}

\def\Label#1{\label{#1}%
  \smash{\hbox to0pt{\raise1ex\hbox{\tiny[#1]}\hss}}}
\def\noLabels{\let\Label=\label}
\def\nobbibitem{\let\bbibitem=\bibitem}

\begin{document}
\noLabels % uncomment for final production
\nobbibitem % uncomment for final production

%\begin{titlepage}

%\begin{flushright}%\vspace{-2cm}
%{\small
%UPR-1154-T  \\ %\vspace{-0.35cm}
%LBNL-60486 \\
%hep-th/0606118}%\\
%\end{flushright}
%\vspace{12 mm}

%\vfil\
%vfil

\begin{center}

\thispagestyle{empty}
{\Large \bf Black Holes as Rubik's Cubes}

\vspace{5mm}

Bart{\l}omiej Czech, 
Klaus Larjo, Moshe Rozali
%\footnote{\tt email:czech, larjo, rozali@phas.ubc.ca} \\

\vspace{5mm}

\bigskip\centerline{\it Department of Physics and
Astronomy}
\smallskip\centerline{\it University of British Columbia}
\smallskip\centerline{\it 6224 Agricultural Road, Vancouver,
BC V6T 1Z1, Canada}
\medskip\centerline{\tt czech, larjo, rozali@phas.ubc.ca}
\bigskip\bigskip\bigskip

%\vfil

\end{center}
\begin{abstract}
\noindent
We propose a unitary toy model of black hole evaporation, in which the entanglement between the interior and exterior degrees of freedom vanishes at late times. Our model possesses the information-free property and satisfies the niceness conditions discussed in the literature. A key feature of the model is that the Hilbert space of black hole internal states contains a vacuum state corresponding to the completely evaporated black hole, which can be reached from any initial state via the Hawking process. Our model suggests a novel quantum cosmological way in which information can get out of an evaporating black hole. 
\end{abstract}
%\vspace{0.5in}

%\end{titlepage}
%\renewcommand{\baselinestretch}{1.05}  %Line spacing
%\tableofcontents %uncomment to display table of contents

\newpage
\setcounter{page}{1}

%\tableofcontents

\section{Introduction}
Black hole evaporation appears to lead to either loss of unitarity or highly entangled remnants  carrying macroscopic entropy \cite{Hawking:1976ra}. A possible resolution of this `information paradox'  is that information about the internal state of the black hole escapes and is available to an observer outside the black hole via the Hawking radiation, or more precisely as small state-dependent corrections on top of the purely thermal radiation.  In \cite{mathurtheorem,mathurmodel} it was argued, subject to certain plausible assumptions, that this cannot the case: neither semiclassical Hawking radiation nor small corrections to it can carry away sufficient amount of information.  Instead, the entanglement entropy between degrees of freedom internal and external to the black hole will increase indefinitely, leading inevitably to either a remnant or to a mixed state, signifying a loss of unitarity.

In this note we present a toy model that satisfies the same set of assumptions, yet avoids the conclusion above: the entanglement entropy turns around after an initial hike and starts decreasing toward zero, which allows the information to escape the black hole before it evaporates completely.  The new ingredient in our toy model is the explicit accounting for pathways or histories whereby the evaporation process can terminate. This is implemented by designating a specific internal state to be `the vacuum' corresponding to an evaporated black hole.  
%This represents a particular small modification of the semi-classical Hawking process. 
While the emitted Hawking radiation is still completely independent of the internal state of the black hole, it may depend on  the mass of the black hole. In our model such dependence takes the simplest possible form: we only stipulate that Hawking radiation ceases once the black hole has evaporated. This suggests a novel mechanism for the way information trapped inside a black hole can be released and made accessible to an outside observer, which we discuss further in the final section.

The plan of this note is as follows. We begin, in Section~\ref{revhorizons}, by providing some background and explaining the conclusions of \cite{mathurtheorem,mathurmodel}. We then proceed to contrast this with our models in Section~\ref{models}. Rather than introduce our complete model from the get go, we start by constructing a few trial toy models which incorporate some, but not all, features that may be expected from models of black hole evaporation. These models all share the property that information becomes accessible to an outside observer before the black hole has completely evaporated. We close Section~\ref{models} with the introduction of our final model, which combines all the desirable characteristics of the trial models of that section. After that, in Section~\ref{information}, we discuss information retrieval and estimate the information retrieval time. We conclude with a discussion of our results and potential directions for future research.

\section{Information-free horizons, niceness conditions and the growth of entanglement entropy}
\label{revhorizons}

Ref.~\cite{mathurmodel} analyzed and contrasted models of black holes evaporation and of burning paper. It found the following crucial difference: while outgoing Hawking radiation is entirely independent of the internal state of the black hole, a burning material completely determines the outgoing particles.\footnote{This difference between black holes and burning paper rules out hypothetical black hole analogues of spodomancy, because escaping Hawking particles are unrelated to the innards of a black hole.} A model in which the outgoing particles are independent of the state of the system is said to have an `information-free horizon'. The information-free property is the first feature we shall demand of the models constructed in this note.

It was argued in \cite{mathurtheorem} that in any unitary black hole model, which has an information-free horizon and satisfies a set of `niceness conditions', the entanglement entropy between the interior and exterior will always increase unboundedly. For our purposes, `niceness conditions' mean that the black hole should have a good semiclassical description, the second feature we shall demand of our models. In \cite{mathurtheorem}, the information-free horizon was implemented by choosing the time evolution to be of the form
\begin{equation}
\psi_i \otimes \chi_i \to  \psi_i \otimes \left( \frac{1}{\sqrt{2}}  |0\rangle_{\rm int}  |0\rangle_{\rm ext} + \frac{1}{\sqrt{2}}|1\rangle_{\rm int}  |1\rangle_{\rm ext} \right) \otimes \chi_i, \label{mathurevolution}
\end{equation}
where $\psi_i$ is the internal state of the black hole, $\chi_i$ is the state of previously emitted Hawking radiation, and the expression in parenthesis represents the state of a newly created Hawking pair at the horizon. The negative energy quantum $| \rangle_{\rm int}$ falls inside the hole and is considered part of $\psi_i$ at the next step, while the other particle $|\rangle_{\rm out}$ escapes and is considered part of $\chi_i$ in future steps. The time evolution entangles the interior and exterior, with the entanglement entropy growing by $\ln 2$ at every step.

\paragraph{A note on small corrections:}  In \cite{mathurtheorem,mathurmodel} small corrections on top of the pair creation (\ref{mathurevolution}) were analyzed in order to probe whether such corrections are sufficient for information retrieval. These corrections were implemented by allowing the pair to be created in a different state with a small probability amplitude  $\epsilon$. It was found that such corrections were not enough to reverse the growth of the entanglement entropy.  Hence, the hope that small state-dependent corrections to Hawking radiation could transmit information about the internal state  was shown to be false.  

As we show below, information retrieval in our models does not rely on such corrections.  Our models can be easily extended to include such corrections and we have checked that the behavior of the entanglement entropy is not strongly affected by such extensions. Therefore, for the sake of simplicity of presentation, we shall only consider the basic Hawking pair creation process.

\section{The models}
\label{models}
We start this section by presenting the physical reasons why it is possible to get around the argument of \cite{mathurtheorem}. We then devise three unitary trial models, which contain some strengths and weaknesses. In the final subsection we combine the trial models to construct the final model of black hole evaporation, which satisfies all the requirements -- unitarity, niceness conditions and the information-free property.

\subsection{The strategy}
\label{strategy}
It is crucial to realize that while the time evolution (\ref{mathurevolution}) is sufficient for an information-free horizon, it is not implied by it.  Hence, we can consider other types of time evolutions.  It is also clear that the evolution (\ref{mathurevolution}) can never lead to the evaporation of the black hole, because at every step a new particle is tensored onto the internal state $ \psi_i$, thus leading to an ever-increasing complexity of the internal state of the black hole. 

 We wish to modify (\ref{mathurevolution}) in such a way that the outgoing particles are still tensored onto the previously escaped Hawking radiation, but the infalling particles act on the black hole microstate as operators, rather than just enlarging the Hilbert space:
\begin{equation}
\psi_i \otimes \chi_i \to \psi_i \, \overleftarrow{S} \otimes \chi_i. \label{ourevolution}
\end{equation}
The left-arrow refers to our typographical convention: the pair is created at the horizon, and the escaping particle moves out (right), while the other particle falls in (left).  Thus $\overleftarrow{S}$ denotes an operator acting on the internal state $\psi_i$.

The physical motivation for this is simple: while the information-free property guarantees that the created pair is independent of the details of the internal microstate, the pair may still depend on one property of the black hole: the mass.  After all, large black holes are colder than small ones and an evaporated black hole ceases to radiate. Hence, black hole states are graded by their mass, which in turn may affect the form of the Hawking radiation. In some of our models we shall not use the whole grading. Instead, we will learn that it is sufficient to demand the existence of a single vacuum state (evaporated black hole), which may be reached from any other state via transitions correlated with randomly generated Hawking radiation. 

\paragraph{An example -- the Rubik's cube:} Think of the interior of the black hole as a Rubik's cube. The interior Hilbert space is spanned by the $4\times 10^{19}$ configurations of the cube. We declare that the solved configuration of the cube is the internal vacuum and corresponds to the black hole having evaporated. When a  fluctuation creates a particle-antiparticle pair at the horizon, the black hole is seen to emit a Hawking particle. The negative energy quantum trapped inside the black hole affects the internal degrees of freedom (cube configurations). We can think of this effect as being enacted by (a linear combination of) the basic Rubik moves, with the proviso that once the cube is solved (once the black hole evaporates), the Hawking process is discontinued. Of course, the way in which the Hawking process affects the internal degrees of freedom is entangled with the escaping particles. Each classical history of the black hole looks like a series of moves which eventually solves the cube. Quantum mechanically, the wavefunction of the black hole gradually concentrates on the internal vacuum, because any random set of moves eventually leads to a solution of the cube. 

One could object that it is a considerable restriction to consider a finite-dimensional internal space, as then by construction the system will eventually concentrate on the vacuum.  However, for a black hole of a given mass the number of microstates is finite and given by the exponential of the entropy. As Hawking radiation can only decrease the mass of a black hole, throughout the evaporation process the system can only access the finitely many states whose mass is no greater than the initial mass of the black hole. We believe that using the Rubik model sufficiently captures this. It is certainly possible to augment the model with infinitely many additional microstates of higher mass without affecting the evolution of the entanglement entropy.

\subsection{Trial models}

\subsubsection{Model~I -- Full randomness}
For the purpose of  simulations, we replace the Rubik model with a simpler analogue, Model~I. Let the internal Hilbert space be spanned by configurations of numbers $1, 2, 3, 4$ arranged in $2 \times 2$ tableaux: 
\begin{equation}
\psi_i \equiv
\begin{array}{|c|c|}
\hline
a & b \\
\hline
c & d \\
\hline
\end{array}
\label{intbasis}
\end{equation}
We take these 24 states to be orthonormal. We single out the state
\begin{equation}
|\rm{vac}\rangle \equiv
\begin{array}{|c|c|}
\hline
1 & 2 \\
\hline
3 & 4 \\
\hline
\end{array}
\label{intvacuum}
\end{equation}
as the internal vacuum, corresponding to the black hole having completely evaporated. We define three elementary operations
\begin{equation}
\begin{array}{|c|c|}
\hline
a & b \\
\hline
c & d \\
\hline
\end{array}  \, \, \, \overleftarrow{L} = \begin{array}{|c|c|}
\hline
c & b \\
\hline
a & d \\
\hline
\end{array}\, \, , \qquad\quad
\begin{array}{|c|c|}
\hline
a & b \\
\hline
c & d \\
\hline
\end{array} \, \, \,  \overleftarrow{R}= \begin{array}{|c|c|}
\hline
a & d \\
\hline
c & b \\
\hline
\end{array} \, \,  , \qquad\quad
 \begin{array}{|c|c|}
\hline
a & b \\
\hline
c & d \\
\hline
\end{array} \, \, \, \overleftarrow{U} = \begin{array}{|c|c|}
\hline
b & a \\
\hline
c & d \\
\hline
\end{array}\, \, \, . \label{deflru}
\end{equation}
One can think of these three operations as moves on a puzzle whose objective is to arrange the four numbers as in expression~(\ref{intvacuum}). We also define a fourth operation $\overleftarrow{N}$, which leaves the state unchanged.\footnote{{\bf N}o move, {\bf L}eft, {\bf R}ight, {\bf U}pper.}

The external states $\chi_i$ are ordered sequences of Hawking-radiated particles. We take the external particles to be of four\footnote{Although $n$ should really be seen as the absence of an emitted particle at that timestep, it is convenient to call it a `particle'. } basic types $\{n,l,r,u\}$, so $\chi_i$ are words built from these four letters, e.g. $unlrl$. All such states are taken to be orthonormal. Two states $\psi_i \otimes \chi_j$ and $\psi_k \otimes \chi_l$ are orthogonal unless the internal states $\psi_i$ and $\psi_k$ and the radiated Hawking words $\chi_j$ and $\chi_l$ both agree. 

To define the time evolution (\ref{ourevolution}) we must choose the operator $\overleftarrow{S}$ enacting the Hawking radiation. When acting on states whose interior component is \emph{not} the vacuum,  the action is chosen to be: 
\begin{equation}
\overleftarrow{S} = \frac{1}{2}\left( \overleftarrow{N} \otimes n + \overleftarrow{L} \otimes l + \overleftarrow{R} \otimes r+ \overleftarrow{U} \otimes u \right). \label{def1}
\end{equation}
In these tensor expressions, the left component acts on the internal state (\ref{intbasis}) according to (\ref{deflru}) while the right component prepends the particle $n,l,r,u$ to the Hawking-radiated word $\chi_i$. An example of one time step would be:
\begin{align}
\begin{array}{|c|c|}
\hline
4 & 2 \\
\hline
1 & 3 \\
\hline
\end{array} \, \,  |unlrl\rangle & \longrightarrow \,\,\,
\begin{array}{|c|c|}
\hline
4 & 2 \\
\hline
1 & 3 \\
\hline
\end{array} \, \, \overleftarrow{S} |unlrl\rangle  \label{exevolution} \\ & =
\frac{1}{2} \!
\left(\,
\begin{array}{|c|c|}
\hline
4 & 2 \\
\hline
1 & 3 \\
\hline
\end{array} \, |nunlrl\rangle
+ 
\begin{array}{|c|c|}
\hline
1 & 2 \\
\hline
4 & 3 \\
\hline
\end{array} \, |lunlrl\rangle
+
\begin{array}{|c|c|}
\hline
4 & 3 \\
\hline
1 & 2 \\
\hline
\end{array} \, |runlrl\rangle
+
\begin{array}{|c|c|}
\hline
2 & 4 \\
\hline
1 & 3 \\
\hline
\end{array} \, |uunlrl\rangle
\!\right)\, . \nonumber
\end{align}
On the other hand, when acting on a state whose internal component is the evaporated state $|{\rm vac} \rangle$, the state is left unchanged, e.g.:
\begin{equation}
\begin{array}{|c|c|}
\hline
1 & 2 \\
\hline
3 & 4 \\
\hline
\end{array} \,\,\overleftarrow{S} \, |unlrl\rangle = \,
\begin{array}{|c|c|}
\hline
1 & 2 \\
\hline
3 & 4 \\
\hline
\end{array} \, |nunlrl\rangle \, .
\label{actonvacuum}
\end{equation}
Recall that  the prepended $n$ should be viewed as the absence of a particle. Eq.~(\ref{actonvacuum}) states that once the black hole evaporates, the Hawking process terminates and the state does not evolve further.

It is easy to check that this model is unitary. It also enjoys the information-free property, since from the viewpoint of an external observer the Hawking process randomly spits out particles $l,r,u$, each of them radiated an equal fraction of the time.  The only exception to this is the vacuum state, in which the Hawking process ceases.

The time evolution eventually forces every internal state to be brought to the internal vacuum. Thus, if one considers the internal degrees of freedom in isolation, their time evolution is not unitary. Physically this is a trivial point, a consequence of the fact that black holes evaporate. However, it highlights a key difference between Model~I and eq.~(\ref{mathurevolution}). In the latter, the dynamics of the internal degrees of freedom is unitary all by itself, without accounting for the exterior subsystem.

\begin{figure}[t]
\begin{center}$
\begin{array}{cc}
\includegraphics[width=0.4\textwidth]{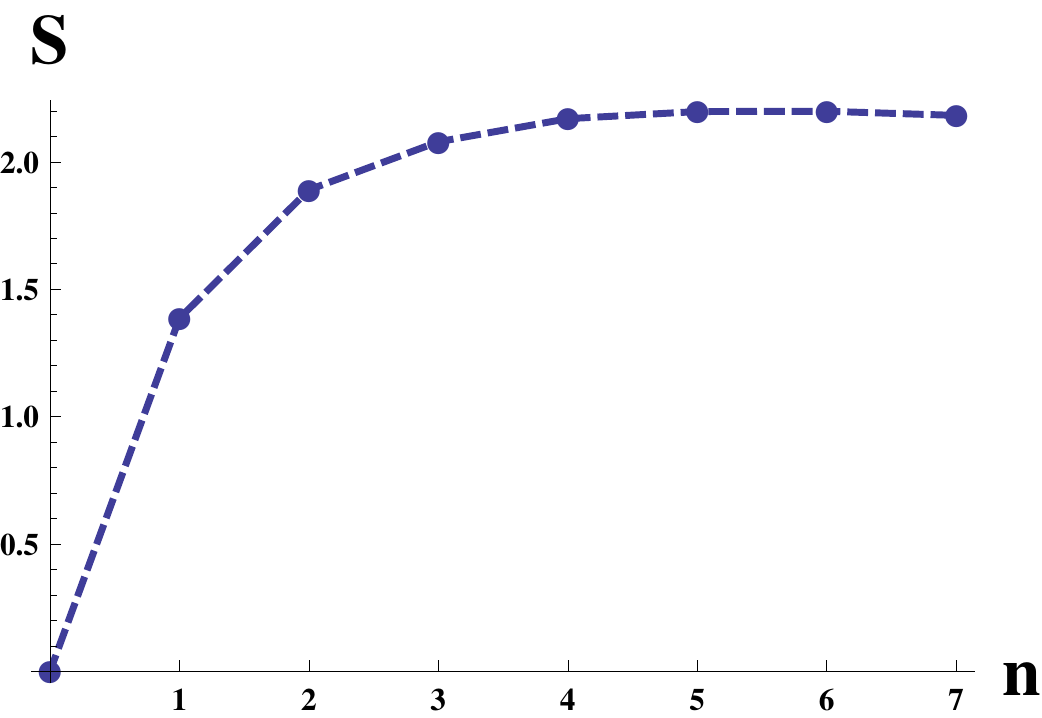} \quad \qquad & 
\includegraphics[width=0.4\textwidth]{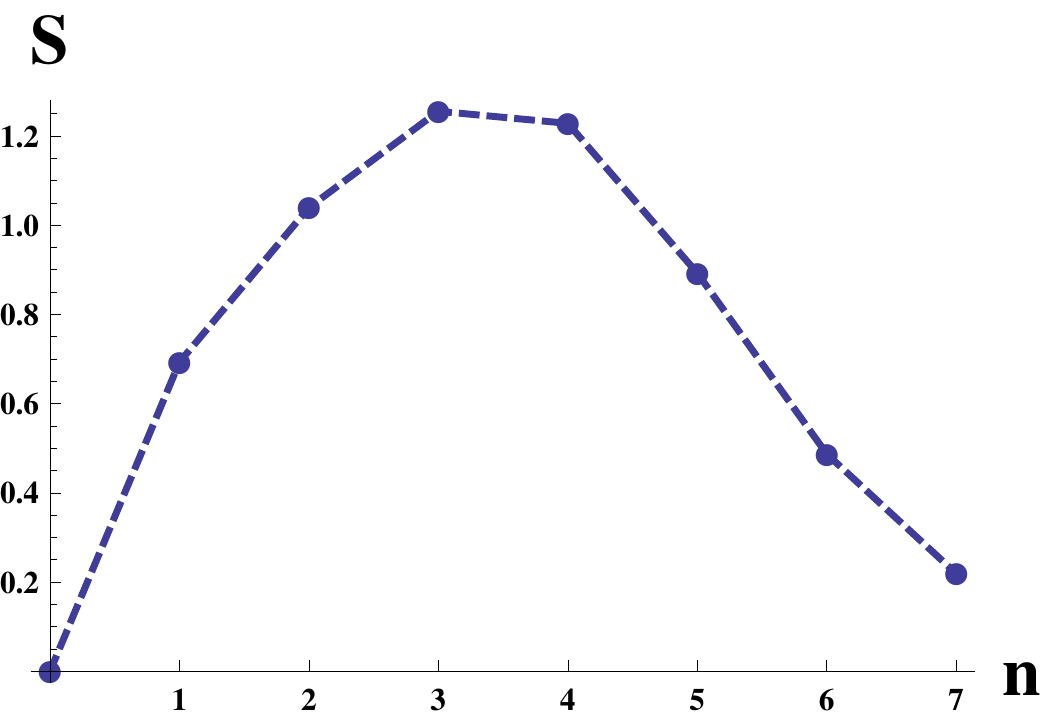}
\end{array}$
\end{center}
\caption{\label{fig-model12} Entanglement entropies as functions of time for Models~I (left) and II (right). In both cases the entropy reaches a maximum and turns around after some time step. The turn-around time and other details of the plots depend on the model and the initial state.}
\end{figure}

Because the black hole eventually evaporates, the entanglement entropy between the interior and exterior degrees of freedom cannot avoid turning around and approaching zero at late times. We have simulated Model~I numerically and the entanglement entropy is plotted in Figure~\ref{fig-model12}.  The initial state for the evolution was chosen to be
$\begin{array}{|c|c|}
\hline
3 & 2 \\
\hline
1 & 4 \\
\hline
\end{array}\,$; other choices yield similar plots.  The location of the turning point depends on how many moves away from the vacuum the initial state was. The model is too simple for deriving quantitative results about black holes. 

\paragraph{Strengths and weaknesses:} 
Model~I incorporates an information-free horizon, but it does not satisfy the niceness conditions.  One way of seeing this is to note that while an arbitrary puzzle is randomly solved in an average of roughly 40 moves, the standard deviation in the number is of the same order of magnitude. Thus, given an initial black hole it is not possible to accurately predict when it evaporates, which contradicts the expected behavior of a large semi-classical black hole.  

A comparison with semiclassical black holes reveals one other deficit. The process of black hole evaporation is a gradient process, characterized by a number of macroscopic parameters -- mass, entropy, inverse temperature -- continually decreasing to zero. Model~I has no viable counterpart to these quantities, because its time evolution draws on randomness and does not respect any grading of the internal Hilbert space. Our next model is designed to fix this deficiency.

\subsubsection{Model~II -- The hidden hand}

We require that the black hole shrink which each particle emitted. To do so, we grade the basis of the internal Hilbert space (\ref{intbasis}) by the minimal number of moves required to solve the puzzle. Then we modify the time evolution of Model~I by the following rule: after applying the operator (\ref{def1}) to an internal state $\psi_i$, we project out those wavefunction components that are further away from the solution of the puzzle than $\psi_i$ and normalize the remaining wavefunction. This rule ensures that the evolution proceeds `toward the vacuum' and defines Model~II. 

As an example, consider again the time evolution (\ref{exevolution}), which now becomes
\begin{multline}
\begin{array}{|c|c|}
\hline
4 & 2 \\
\hline
1 & 3 \\
\hline
\end{array} \, \,  |unlrl\rangle  \longrightarrow \,\,\, \\
\cancel{\frac{1}{\sqrt{4}}} 
\frac{1}{\sqrt{3}}\!
\left(\,
\begin{array}{|c|c|}
\hline
4 & 2 \\
\hline
1 & 3 \\
\hline
\end{array} \, |nunlrl\rangle
+ 
\cancel{\begin{array}{|c|c|}
\hline
1 & 2 \\
\hline
4 & 3 \\
\hline
\end{array} \, |lunlrl\rangle}
+
\begin{array}{|c|c|}
\hline
4 & 3 \\
\hline
1 & 2 \\
\hline
\end{array} \, |runlrl\rangle
+
\begin{array}{|c|c|}
\hline
2 & 4 \\
\hline
1 & 3 \\
\hline
\end{array} \, |uunlrl\rangle
\!\right).
\label{model2ex}
\end{multline}
The initial state is four moves from the vacuum, and the second component is projected out because it is five moves from the vacuum; retaining it would be akin to the black hole expelling a negative energy Hawking particle.  Figure \ref{fig-model12}  plots the entanglement entropy for Model~II with the initial state $\begin{array}{|c|c|}
\hline
3 & 1 \\
\hline
4 & 2 \\
\hline
\end{array}\,$.  The entropy starts decreasing sooner for Model~II, because the evolution has access to fewer internal states than in Model~I.\footnote{For computational reasons we chose different initial states for the two plots; hence one should not try to compare them too closely.}

\paragraph{Strengths and weaknesses:} 
Model~II is unitary and has the desirable feature that there is a quantity (distance from vacuum), which decreases in the process of black hole evaporation. An analogue of Model~II, in which the $2 \times 2$ tableaux are replaced with more complicated puzzles to produce longer black hole lifetimes, will satisfy the niceness conditions. To see this, note that the lifetime of a Model~II black hole is roughly given by the negative binomial distribution \cite{nbd} with parameters $R,P$ set to the distance of the initial state to the vacuum and the probability of emitting the particle $n$ at each evolution step, respectively. The standard deviation-to-mean ratio of the negative binomial distribution is $\sqrt{P/R}$, but $P \leq 1/2$  and $R$ is of the same order of magnitude as the lifetime of the black hole. 

However, Model~II does not have an information-free horizon, because its Hawking process depends on the state of the black hole. Indeed, an outside observer obtains information about the internal state by observing the absence of certain particles in the radiation signal, such as particle $l$ in example (\ref{model2ex}).

\subsubsection{Model~III -- The depository}
We have now achieved both the niceness conditions and an information-free horizon, though not in the same model.  Before combining the positive features of Models~I and II, let us address another potential objection to time evolution (\ref{ourevolution}). Assuming that the internal degrees of freedom of a black hole are in some way localized, either in the deep interior or spread across the horizon, one may be wary that a Hawking emission has an immediate, global effect on the internal state. In order to address this we introduce a depository of particles, into which infalling particles are placed until they have fallen far enough to affect the internal state. Hence, the states are of the form 
\begin{equation}
\begin{array}{|c|c|}
\hline
4 & 2 \\
\hline
1 & 3 \\
\hline
\end{array} \, \{ \overleftarrow{L} \overleftarrow{N} \overleftarrow{U} \} \,  |unlrl\rangle,
\end{equation}
where the length of the depository is taken to be three particles.  Each time a pair is created, the infalling particle is placed at the right end of the depository (the horizon). The other particles in the depository are then moved left by one unit and the leftmost particle ($\overleftarrow{L}$ in the example above) operates on the interior state.

The entanglement entropy in this model again increases and decreases.  It is perhaps noteworthy that for the first $p$ steps, $p$ being the length of the depository, the entropy increases linearly as in the models of \cite{mathurtheorem,mathurmodel}. Hence those models can be interpreted as a special case of Model~III with the length of the depository taken to infinity.

Model~III is unitary, but it inherits the deficiences of its predecessors. It violates the niceness conditions or lacks an information-free horizon, depending on whether one adds the depository to Model~I or II. 

\subsection{The Final Model}
\label{finalmodel}
We now define a model that incorporates the positive aspects of Models~I and II without sharing their weaknesses. For simplicity we refrain from including the depository of Model~III; including it would be trivial.  

\begin{figure}[t]
\begin{center}
\includegraphics[width=0.45\textwidth]{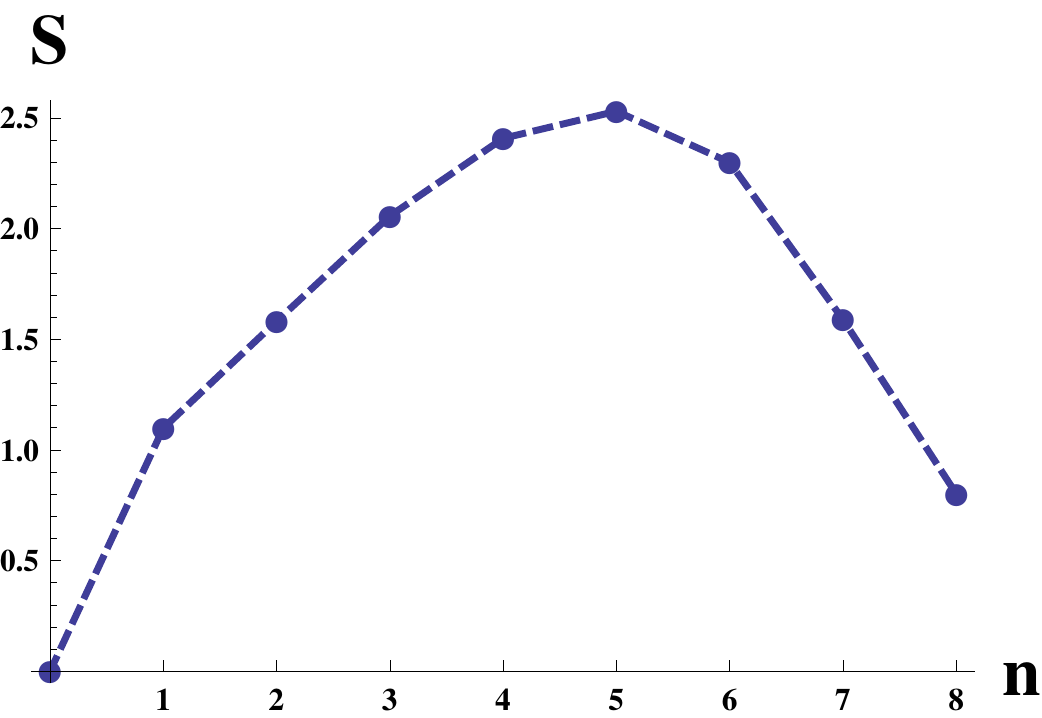} 
\end{center}
\caption{\label{fig-modelfinal}  Entanglement entropy versus time in the Final Model. For computational reasons, this plot was obtained in a simplified version of the model, in which Rubik's cubes or square tableaux were replaced with copies of a simpler puzzle with only six discrete configurations.}
\end{figure}

Consider a model in which the interior of the black hole consists of a large number $E$ of unsolved Rubik's cubes or puzzles of the type considered in Model~I. The quantity $E$ will decrease in the process of black hole evaporation, analogous to energy, entropy or inverse temperature; in what follows we refer to this quantity as `energy' in inverted commas. The basis states are of the form
\begin{equation}
\overbrace{ \left\{ \begin{array}{|c|c|}
\hline
4 & 2 \\
\hline
1 & 3 \\
\hline
\end{array}\,\,  , \ldots, \,\, \begin{array}{|c|c|}
\hline
2 & 4 \\
\hline
3 & 1 \\
\hline
\end{array} \right\}}^E  \,   |unlrl\rangle.
\label{exampstate}
\end{equation}
The time evolution takes the form familiar from Model~I except that the operators $\overleftarrow{N},\overleftarrow{L},\overleftarrow{R},\overleftarrow{U}$ act on each tableau individually. In order to guarantee unitarity, we further stipulate that each time a square is solved (brought to the vacuum state), an extra particle $q$ is emitted. Augmenting our model with the $q$-particle is physically justified, because an outside observer ought to be able to detect that the black hole has lost `energy'. The following example illustrates the time evolution involving a $q$-particle:
\begin{multline}
\left\{ 
\begin{array}{|c|c|}
\hline
4 & 2 \\
\hline
1 & 3 \\
\hline
\end{array}
\, , \,
\begin{array}{|c|c|}
\hline
2 & 1 \\
\hline
3 & 4 \\
\hline
\end{array} 
\, \right\}\!
|\rangle  \longrightarrow \\
\frac{1}{2}\!\left(
\left\{ 
\begin{array}{|c|c|}
\hline
4 & 2 \\
\hline
1 & 3 \\
\hline
\end{array}
\, , \,
\begin{array}{|c|c|}
\hline
2 & 1 \\
\hline
3 & 4 \\
\hline
\end{array} 
\, \right\}\!  
|n\rangle
+
\left\{ 
\begin{array}{|c|c|}
\hline
1 & 2 \\
\hline
4 & 3 \\
\hline
\end{array}
\, , \,
\begin{array}{|c|c|}
\hline
3 & 1 \\
\hline
2 & 4 \\
\hline
\end{array} 
\, \right\}\!  
|l\rangle
+
\left\{ 
\begin{array}{|c|c|}
\hline
4 & 3 \\
\hline
1 & 2 \\
\hline
\end{array}
\, , \,
\begin{array}{|c|c|}
\hline
2 & 4 \\
\hline
3 & 1 \\
\hline
\end{array} 
\, \right\}\!  
|r\rangle
+
\begin{array}{|c|c|}
\hline
2 & 4 \\
\hline
1 & 3 \\
\hline
\end{array}\,\,
|qu\rangle
\right)
\label{finalmodelex}
\end{multline}
In the wavefunction component in the last term the second square is erased, because the transposition $\overleftarrow{U}$ solved that puzzle. An outside observer can detect the drop in `energy' by seeing the outgoing particle $q$. The particle $q$ is necessary to preserve unitarity; if it were absent, the state obtained after one time step from $\begin{array}{|c|c|}
\hline
4 & 2 \\
\hline
1 & 3 \\
\hline
\end{array}\,\,|\rangle$ would have non-trivial overlap with (\ref{finalmodelex}). A sample plot of the entanglement entropy in this model is presented in Figure~\ref{fig-modelfinal}.

\paragraph{Properties and interpretation:}
The model is unitary; the $q$-particles ensure unitarity in all cases that do not follow directly from Model~I. Our model also has the desired gradient property that was missing in Model~I, because the number of unsolved puzzles $E$ can only decrease in the course of the evolution.

The other two desirable characteristics, an information-free horizon and the niceness conditions, require some explanation. Because the production of particles $n,l,r,u$ follows Model~I, it is by itself information-free, but $q$-particles inform the outside observer about the count of `energy' lost in each time step. However, recall that the rationale for the information-free property is the semiclassical picture of black holes, in which only \emph{internal} degrees of freedom must remain invisible to outside observers. In contrast, \emph{global} properties such as mass or temperature can and should be visible to outside observers and affect the form of Hawking radiation. Since the Final Model was motivated by a desire to construct an evaporation-decreasing, global quantity analogous to energy, it is not surprising that that quantity becomes accessible to the outside observer as the count of outgoing $q$-particles. It is an interesting lesson that in our model that access is exacted by unitarity.

In order to establish the niceness conditions, we will require the system to be large in a certain quantifiable sense. As a first observation, notice that the number of $q$-particles emitted by our black hole as a function of time is determined by a random walk in the space of internal states, which is generated by the basic moves (definition~(\ref{deflru}) in our current model). Whenever the random walk takes one of the $E$ initial component puzzles to the vacuum, a $q$-particle is emitted. But when the random walk self-intersects, no $q$-particles are produced, because the puzzles which would be solved on that step had been solved and erased from the state description before. We shall now argue that in the regime of high $E$, the niceness conditions hold whenever the internal space is large and rich enough so that random walks typically do not self-intersect for long times. This, of course, does not hold for the $2 \times 2$ puzzles used in eqs.~(\ref{exampstate}-\ref{finalmodelex}), but it does hold for Rubik's cubes. 

Take $E$ to be much larger than the size of the internal space $K$ (in the $2 \times 2$ model $K=24$ while in the Rubik's cube $K=4\times 10^{19}$), so that almost all internal states contain unsolved puzzles in almost all configurations. Initially, the outside observer will see at every time step approximately the same number of $q$-particles, $E/K$, with typical relative deviation of order $\sqrt{K/E}$. The niceness conditions hold firmly so long as these deviations remain small, that is so long as random walks do not (frequently) self-intersect. When self-intersections begin to kick in, the wavefunction of the outgoing radiation develops components marked by conspicuous absences of $q$-particles. At this stage, the nice semiclassical description of the state is still maintained in a coarse-grained sense: the average number of detected $q$-particles per $k$ steps is well-behaved for sufficiently large $k$. But when the random walk has covered almost all the internal space, leaving only a few unsolved puzzles scattered over distant regions of the internal space, then large uncertainties take over and the niceness conditions are gone. This is easily interpreted: at advanced stages of evaporation, black holes are nearly Planck-sized and not well-described by semiclassical physics.

At last, we remark that our model is too simple to reproduce the evaporation of black holes quantitatively. In particular, in our model large black holes evaporate faster while small black holes linger for longer times, the opposite of the standard relation $T \propto M^{-1}$.

\section{Information retrieval}
\label{information}
It is of interest to analyze how information implicit in the initial state of a black hole escapes from a system with an `information-free' horizon with the emitted Hawking radiation. To analyze this quantitatively we follow the discussion of \cite{hp}, which quantified the issue of retrieval of quantum information from black holes. 

We take the initial state of the black hole to be entangled with a reference system held by Charlie. Hence the quantum state lives in the Hilbert space 
\begin{equation}
\mathcal{H}_{\rm int} \otimes \mathcal{H}_{\rm ext} \otimes \mathcal{H}_{\rm Charlie},
\end{equation}
where time evolution on $\mathcal{H}_{\rm Charlie}$ is taken to be trivial. For simplicity of presentation we work with the black holes of Model~I, but the discussion applies equally well to our Final Model.

Clearly, one cannot demand Bob, the outside observer, to determine the exact quantum state of the radiation or the initial state of the black hole. Rather, we say that the information about the initial state has come out with accuracy $(1-\delta)$ when any measurement that Charlie can make on the reference system with outcome probabilities $\{p_i\}$ can be reproduced by Bob with outcome probabilities $\{p_i \pm \mathcal{O}(\delta)\}$. We know that this will eventually happen, because the interior wavefunction concentrates in time on the vacuum configuration so that we forget nothing by tracing over $\mathcal{H}_{\rm int}$. By unitarity, if the initial entanglement between Charlie and the interior is lost, a compensating entanglement between Charlie and Bob must have formed.

We illustrate this with the parity qubit,\footnote{By qubit we mean a single property of the system, an answer to a yes / no question. It is not implied that the system is necessarily a tensor product of localized binary degrees of freedom.} which in a sense explained below is the hardest observable to decode for Bob. Parity distinguishes even configurations (internal squares that require an even number of transpositions to be solved) from odd ones. A suitable reference system for this qubit is $\mathcal{H}_{\rm Charlie} = {\rm Span}\{ |{\rm even} \rangle, \,\, |{\rm odd} \rangle \}$, with the initial state of the form
\begin{equation}
\psi(t=0) = \sum_{(\,\,)_i {\rm even}} a_i |(\,\,)_i \rangle_{\rm int} \otimes |\rangle_{\rm ext} \otimes |{\rm even}\rangle_{\rm Charlie} + \sum_{(\,\,)_i {\rm odd}} b_i |(\,\,)_i \rangle_{\rm int} \otimes |\rangle_{\rm ext} \otimes |{\rm odd}\rangle_{\rm Charlie},
\end{equation}
where the sums are over even and odd basis configurations in $\mathcal{H}_{\rm int}$. Charlie can measure the parity of the reference system, finding `even' and `odd' with probabilities
\begin{equation}
p^{\rm C}_{\rm even} = \sum_i |a_i|^2 \quad {\rm and} \quad p^{\rm C}_{\rm odd} = \sum_i |b_i|^2.
\end{equation}
In order to quantify how fast Bob can learn about the parity of the black hole, we must understand how to translate his measurements into decisions about parity. This is because in contrast to Charlie, who measures properties of black hole microstates directly, the strings of Hawking particles measured by Bob correspond to {\it paths} between microstates. Bob can reason as follows: because for late times the internal state concentrates on the vacuum, the parity of the string of Hawking particles (whether the string consists of an even or odd number of particles) should on average reflect the parity of the initial state with increasing accuracy. This reasoning, which extends the definition of the qubit of interest to strings of outgoing radiation, is the most tenuous for parity, because on any string that has not yet brought the initial state to the vacuum Bob incurs a $50\%$ chance of mistake. It is in this sense that parity is the hardest qubit for decoding; qubits with lower entropy or more workable correlations give Bob extra headway. 

\begin{figure}[t]
\begin{center}$
\begin{array}{cc}
\includegraphics[width=0.5\textwidth]{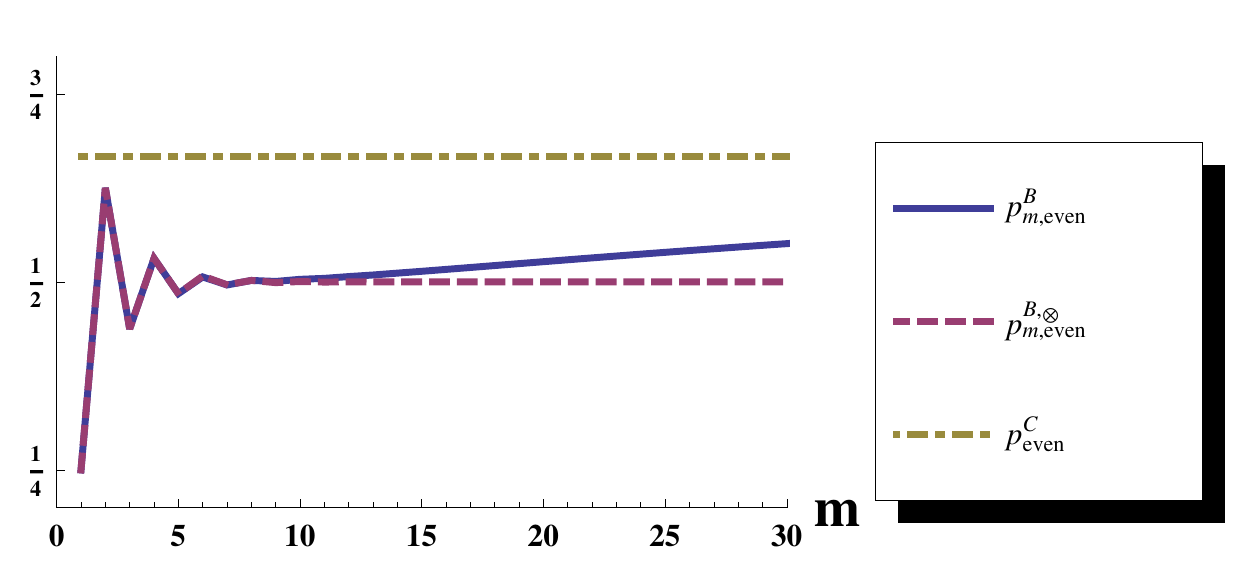} \quad \qquad &
\includegraphics[width=0.4\textwidth]{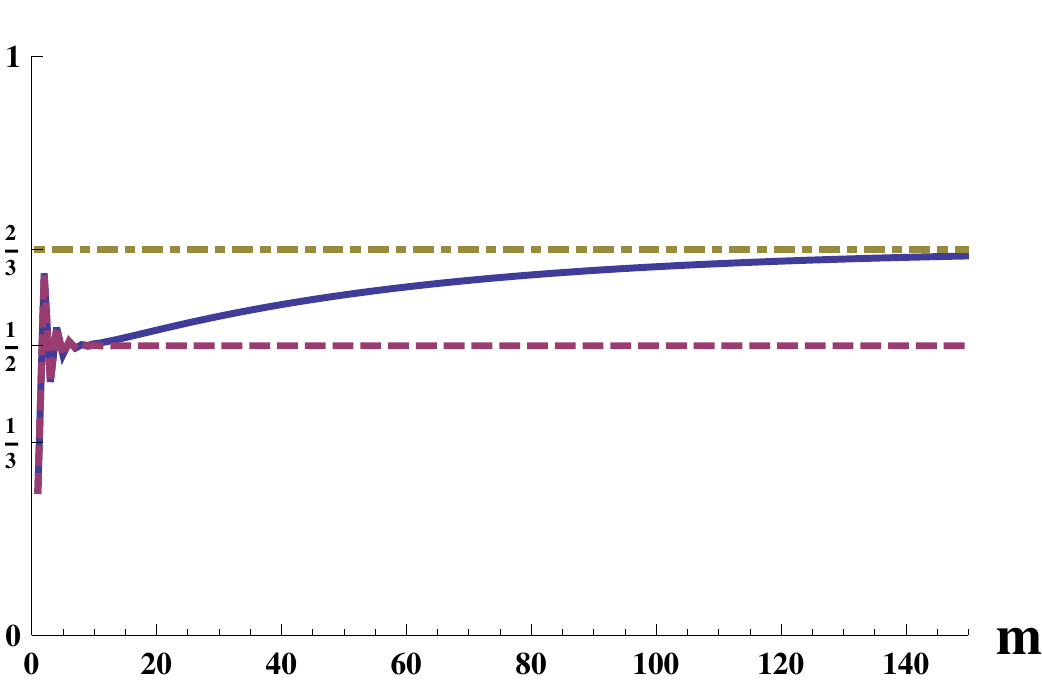}
%\multicolumn{2}{c}{\includegraphics[width=0.5\textwidth]{leadplot.pdf}}
\end{array}$
\end{center}
\caption{\label{fig-inf} $p^{\rm B}_{m,\rm even}$ of state (\ref{exinitial}) starts out equal to $p^{{\rm B},\otimes}_{m,{\rm even}}$, but approaches $p^{\rm C}_{\rm even}=2/3$.  }
%\caption{\label{fig-inf} Top-left: $p^{\rm B}_{m,{\rm even}}$ (solid), $p^{{\rm B},\otimes}_{m,{\rm even}}$ (dashed) and $p^{\rm A}_{\rm even}$ (dash-dot) as a function of time, up to $t=200$.   Top-right: $p^{\rm B}_{m,{\rm even}}$ and $p^{{\rm B},\otimes}_{m,{\rm even}}$ for early times, up to $t=20$.  Bottom: $| p^{\rm B}_{m,\rm even}-  p^{\rm A}_{\rm even}|\cdot (\lambda_{\rm max}/4)^{-m}$ as a function of time.}
\end{figure}

After $m$ steps the state can be written as 
\begin{align}
\psi(t=m) &= \sum_{(\alpha) \,\, {\rm even}} \tilde{a}_{\alpha} |{\rm vac}\rangle_{\rm int} \otimes |(\alpha) \rangle_{\rm ext}\otimes |{\rm even} \rangle_{\rm Charlie} \nonumber \\ &+ \sum_{(\alpha) \,\, {\rm odd}} \tilde{b}_{\alpha} |{\rm vac}\rangle_{\rm int} \otimes |(\alpha) \rangle_{\rm ext}\otimes |{\rm odd} \rangle_{\rm Charlie} + \epsilon_m ({\rm remainder}),
\label{remaindereq}
\end{align}
where $(\alpha)$ denotes strings of $m$ particles and (remainder) denotes the part of the quantum state where the internal microstate has not yet reached the vacuum, with $\epsilon_m \to 0$ as $m\to \infty$. By construction, Bob recovers even and odd radiation strings with probabilities 
\begin{align}
p^{\rm B}_{m,{\rm even}} &=  \sum_\alpha |\tilde{a}_{\alpha}|^2 + \mathcal{O}(\epsilon_m^2) \stackrel{m\to \infty}{\longrightarrow} \sum_i |a_i|^2 = p^{\rm C}_{\rm even}, \\
p^{\rm B}_{m,{\rm odd}} &= \sum_\alpha |\tilde{b}_{\alpha}|^2 + \mathcal{O}(\epsilon_m^2) \stackrel{m\to \infty}{\longrightarrow} \sum_i |b_i|^2 = p^{\rm C}_{\rm odd}.
\end{align}
As an illustration, start with the initial state
\begin{equation}
\psi(t=0) =\sqrt{ \frac{1}{3}} \,\,
\begin{array}{|c|c|}
\hline
4 & 1 \\
\hline
2 & 3\\
\hline
\end{array} \otimes  |{\rm odd}\rangle + \sqrt{\frac{2}{3}} \,\,
\begin{array}{|c|c|}
\hline
2 & 1 \\
\hline
4 & 3 \\
\hline
\end{array} \otimes |{\rm even}\rangle,
\label{exinitial}
\end{equation}
whose components are five and six moves away from the vacuum, respectively. Figure~\ref{fig-inf} plots $p^{\rm B}_{m,\rm even}$. It approaches Charlie's measurement $p^{\rm C}_{\rm even}=2/3$ at roughly $m \sim \mathcal{O}(100)$. This time scale, which is generic for most choices of qubit and initial states, can be recovered quantitatively with the following argument. The Model~I time evolution (\ref{exevolution}-\ref{actonvacuum}) can be thought of as generating paths in the Cayley graph of the permutation group $S_4$ in the presentation generated by $\overleftarrow{L},\overleftarrow{R},\overleftarrow{U}$. To estimate the remainder term in eq.~(\ref{remaindereq}) we must evaluate the fraction of walks of length $m$ that do not visit the vacuum. This is easily done by constructing the adjacency matrix of the Cayley graph\footnote{in a slightly modified form: to account for the moves $\overleftarrow{N}$ one adds to the Cayley matrix the identity matrix and to account for the `sink' in the vacuum configuration one removes its corresponding row and column.} and raising it to power $m$. The largest eigenvalue of that matrix divided by 4 (because a random walk of length $m-1$ produces 4 offspring walks of length $m$) determines the rate at which $p^{\rm B}_{m,\rm even}$ approaches $p^{\rm C}_{\rm even}$. In Model~I, this number turns out to be $0.98$, so that
\begin{equation}
|p^{\rm B}_{m,\rm even} - p^{\rm C}_{\rm even}| \propto 0.98^m , \label{therate}
\end{equation}
where the proportionality constant depends on the choice of qubit and initial state. Thus, Bob's measurements reproduce those of Charlie's with accuracy $\delta$ after $\ln{\delta} / \ln{0.98} \approx 52\ln{\delta^{-1}}$ steps. We did not attempt to evaluate the analogue of (\ref{therate}) for a model based on the full Rubik's cube.

It is interesting to contrast $p^{{\rm B}}_{m,{\rm even}}$ with the analogous quantity in the original model (\ref{mathurevolution}), which we call $p^{{\rm B},\otimes}_{m,{\rm even}}$. 
In that case the likelihood of emitting any particle is $1/4$ at all times. Since the parity of a string is given by the number of non-trivial moves, simple combinatorics gives
\begin{equation}
p^{{\rm B},\otimes}_{m,{\rm even}} = \frac{1}{{4^m}} \sum_{j=0}^{[m/2]} 3^{2j}{\binom{m}{2j}} \stackrel{m\to \infty}{\longrightarrow} \frac{1}{2}.
\end{equation}
We note from the figure that $p^{\rm B}_{m,\rm even}$ and $p^{{\rm B},\otimes}_{m,{\rm even}}$ agree for the first five steps, that is until the wavefunction first develops an internal vacuum component. At that point, $p^{\rm B}_{m,\rm even}$ starts differing from $p^{{\rm B},\otimes}_{m,\rm even}$, because the vacuum no longer emits particles, which favors the production of $n$ over $\{l,r,u\}$. This is the point when information begins to leak out.

The rate (\ref{therate}) should be contrasted with the `mirroring' behavior of \cite{hp}, where it was found that the information thrown into a black hole escapes almost immediately with Hawking radiation.  The difference arises because the outside observer in \cite{hp} is assumed to have been observing the black hole for its whole history and, as a result, be fully entangled with the black hole. In our case Bob has no initial knowledge of the history of the black hole and has to start his measurements from the beginning.

We also note that the information recovery time scale is longer than the time scale discussed by Page \cite{Page:1993df}. This is not unexpected since we are asking a slightly different question: while Page's time scale is roughly the time when an external observer starts distiguishing the Hawking radiation from completely scrambled pure thermal radiation, we are asking for the external observer to be able to reproduce the initial state with a given precision. 
%Thus our time scale is expected to be longer, which is the result we find in simulations.

\section{Discussion}
\label{discussion}

We considered in this note a class of models for black hole evaporation which evade the conclusions of \cite{mathurtheorem}. The new ingredient that causes the entanglement entropy in our models to eventually decrease to zero is that the black hole evaporates rather than growing in complexity ad infinitum. Encoding this requires one to account for the way the negative energy quantum produced in the Hawking process acts on the interal degrees of freedom of the black hole. 

Models~I and II achieve this goal in two different ways. In Model~I, we consider a finite dimensional internal Hilbert space, which guarantees that any random sequence of moves eventually hits the vacuum. This is somewhat unsatisfactory, because one would like to find some quantity that monotonously decreases in the course of the evaporation process, analogously to the mass or horizon area of real black holes. In Model~II black holes evaporate for a different reason: the time evolution is engineered to decrease a certain quantity until the state hits the vacuum. This meets the objection to Model~I and in principle allows one to consider infinite dimensional Hilbert spaces. However, the model does not possess an information-free horizon. In the end, our Final Model combines the attractive features of both predecessors: it satisfies the niceness conditions and has an information-free horizon.

As the details of the Hawking process in our model are independent of the internal state and the outside observer detects the same flux of radiated particles regardless of what sits inside the black hole, how does information get out? 
The mechanism is black hole evaporation. In our models, an outside observer can detect one non-trivial thing about the black hole, namely that it has ceased to emit radiation because it has evaporated. In time, the black hole wavefunction peaks more and more strongly on the evaporated configuration. An outside observer identifies the internal microstate based on the full wavefunction, which is a superposition of different Hawking emission products. In a sense, the information available to an outside observer is a weighted average of the different lifespans of the black hole. To our knowledge, this mechanism for preserving unitarity has not been discussed before in the literature. 

The prominent role that superpositions of internal states play in this mechanism is not too surprising, because any solution to the information paradox must make use of quantum gravity effects (see \cite{review} for a recent survey of the relevant arguments).
The idea of tracing properties of black holes to their being quantum superposition states was explored in \cite{superp}, while the program of understanding black holes through course graining and statistical properties of underlying ensembles of microstates was initiated in \cite{babel}. One challenge that may be raised against our model is that over time the putative black holes may lose their good semiclassical descriptions before they evaporate. On the other hand, it is not clear why quantum gravity should be expected to maintain a good semiclassical description of an evaporating black hole toward the end of the evaporation process. A situation in which quantum fuzziness eventually blurs away a good semiclassical description of a macroscopic object may seem exotic, but perhaps not more so than the black hole information problem itself.

\section*{Acknowledgements}
We thank Philip Argyres, Joshua Davis, Patrick Hayden, Joanna Karczmarek, Thomas Levi and Mark Van Raamsdonk for discussions. We are supported in part by Natural Sciences and Engineering Research Council of Canada and KL is also supported by the Institute of Particle Physics.

\end{document}